%% file: main.tex
\documentclass[a4paper,10pt]{IEEEtran}
\renewcommand{\baselinestretch}{0.99}
\usepackage{pgfplots}
\usetikzlibrary{shapes.multipart,intersections}
\usepackage{cite}
\usepackage{amsmath,amssymb,amsfonts,steinmetz}
\usepackage{algorithmic}
\usepackage{graphicx}
\usepackage{textcomp}
\usepackage{xcolor,flushend}
\def\BibTeX{{\rm B\kern-.05em{\sc i\kern-.025em b}\kern-.08em
    T\kern-.1667em\lower.7ex\hbox{E}\kern-.125emX}}

\usepackage{tikz}
\usetikzlibrary{calc}
\makeatletter
\newcommand{\gettikzxy}[3]{%
  \tikz@scan@one@point\pgfutil@firstofone#1\relax
  \edef#2{\the\pgf@x}%
  \edef#3{\the\pgf@y}%
}
\makeatother
\usepackage[T1]{fontenc}
\usepackage[latin9]{inputenc}
\usepackage{bm}
\usepackage{amsmath}
\usepackage{amssymb}
\usepackage{babel}
\usepackage{comment}
\usepackage{subcaption}
\usepackage{graphicx}
\usepackage{algorithm,algorithmic}
\usepackage{xcolor}
\usepackage{gensymb}
\usepackage{cite}
\usepackage{enumitem}
\usepackage{url}

\input{./Definitions.tex}

\usepackage{amsthm}

\pgfplotsset{compat=1.16}

\begin{document}
\title{Communications-Centric Secure ISAC\\ with Hybrid Reconfigurable Intelligent Surfaces}
\author{\IEEEauthorblockN{
Ioannis Gavras and George C. Alexandropoulos 
} 
\\
\IEEEauthorblockA{Department of Informatics and Telecommunications, National and Kapodistrian University of Athens\\Panepistimiopolis Ilissia, 16122 Athens, Greece}
\\
\IEEEauthorblockA{emails: \{giannisgav, alexandg\}@di.uoa.gr}\vspace{-0.4cm}
}

\maketitle
\begin{abstract}
Hybrid reconfigurable intelligent surfaces (HRISs) constitute an emerging paradigm of metasurfaces that empowers the concept of smart wireless environments, inherently supporting simultaneously communications and sensing. Very recently, some preliminary HRIS designs for Integrated Sensing And Communications (ISAC) have appeared, however, secure ISAC schemes are still lacking. In this paper, we present a novel communications-centric secure ISAC framework capitalizing on the dual-functional capability of HRISs to realize bistatic sensing simultaneously with secure downlink communications. In particular, we jointly optimize the BS precoding vector and the HRIS reflection and analog combining configurations to enable simultaneous accurate estimation of both a legitimate user and an eavesdropper, while guaranteeing a predefined threshold for the secrecy spectral efficiency, with both operations focused within an area of interest. The presented simulation results validate the effectiveness of the proposed secure ISAC design, highlighting the interplay among key system design parameters as well as quantifying the trade-offs between the HRIS's absorption and reflection coeffcients.
\end{abstract}

 

\let\thefootnote\relax\footnotetext{This work was supported by the SNS JU projects TERRAMETA and 6G-DISAC under the EU's Horizon Europe research and innovation programme under Grant Agreement numbers 101097101 and 101139130, respectively. 
}

\pagenumbering{gobble}

\section{Introduction}

The emerging paradigm of Integrated Sensing And Communications (ISAC)\cite{mishra2019toward} has garnered increasing attention across both research and standardization fronts, being regarded as the cornerstone of the upcoming sixth Generation (6G) of wireless networks, advancing on the evolving use cases of its predecessor~\cite{6G-DISAC_mag,stylianopoulos2025distributed}. A key aspect of ISAC is the concept of smart radio environments, primarily enabled by the technology of Reconfigurable Intelligent Surfaces (RISs) \cite{RIS_ISAC,18}. These surfaces enable programmable control over the environment, making it possible to realize advanced functionalities such as simultaneous localization and environmental radio mapping \cite{kim2023ris}, both of which are essential to the ISAC framework.

The concept of Hybrid Reconfigurable Intelligent Surfaces (HRISs) has recently emerged as a novel paradigm that evolves traditional RIS architectures by integrating both reflective and absorptive functionalities within each of their meta-atoms. By embedding power splitters within each unit cell~\cite{hybrid_meta-atom}, a portion of the incoming signal is absorbed and routed to dedicated Radio Frequency (RF) reception chains. This design enables HRISs to support self-reconfiguration\cite{alexandropoulos2023hybrid} and localization capabilities~\cite{alexandropoulos2020hardware,ghazalian2024joint}, ultimately establishing them as a key enabler for ISAC systems. Beyond traditional user positioning, recent studies have demonstrated that the additional beamforming degrees of freedom provided by intelligent metasurfaces can significantly enhance large-scale sensing performance~\cite{alexandropoulos2023hybrid,gavras2024simultaneous}, offering significant potential for area-wide sensing applications and scenarios where target positioning uncertainties need to be taken under consideration.

In this paper, we present a novel secure ISAC framework leveraging the simultaneous tunable reflection and sensing capability of HRISs. In particular, an HRIS is jointly optimized with the BS digital beamforming to realize a bistatic sensing system simultaneously with a metasurface-empowered secure DownLink (DL) system. We derive the Cram\'{e}r-Rao Bound (CRB) for the position estimation of both legitimate User Equipment and an Eavesdropper (Eve), which is then deployed in the objective of a novel secrecy-rate-constrained ISAC problem formulation including the BS precoder and the HRIS dual-functional configuration as the design parameters. Our numerical evaluations underscore the effectiveness of the proposed secure ISAC design, even in the presence of position uncertainties for the UE and Eve. In addition, the role of the HRIS absorption coefficient on the trade-off between sensing and secure DL communications is quantified.

\textit{Notations:}
Vectors and matrices are represented by boldface lowercase and uppercase letters, respectively. The transpose, Hermitian transpose, and inverse of $\mathbf{A}$ are denoted as $\mathbf{A}^{\rm T}$, $\mathbf{A}^{\rm H}$, and $\mathbf{A}^{-1}$, respectively. $\mathbf{I}_{n}$, $\mathbf{0}_{n}$, and $\boldsymbol{1}_n$ ($n\geq2$) are the $n\times n$ identity and zeros' matrices and the ones' column vector, respectively. $[\mathbf{A}]_{i,j}$ is $\mathbf{A}$'s $(i,j)$-th element, $\|\mathbf{A}\|$ gives its Euclidean norm, and $\text{Tr}\{\mathbf{A}\}$ its trace. $|a|$ ($\Re\{a\}$) returns the amplitude (real part) of complex scalar $a$. $\mathbb{C}$ is the complex number set, $\jmath\triangleq\sqrt{-1}$ is the imaginary unit, $\mathbb{E}\{\cdot\}$ is the expectation operator, and $\mathbf{x}\sim\mathcal{CN}(\mathbf{a},\mathbf{A})$ indicates a complex Gaussian 
vector with mean $\mathbf{a}$ and covariance matrix $\mathbf{A}$. 

\section{System and Channel Models}\label{Sec: System_Sec}
\begin{figure}[!t]
	\begin{center}
    \includegraphics[scale=0.7]{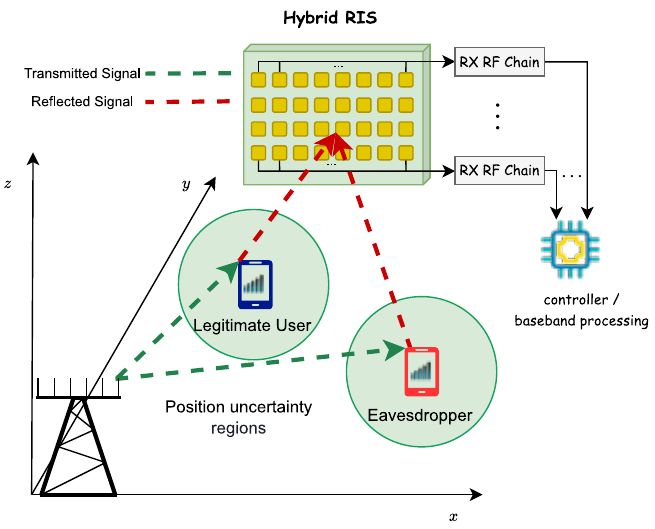}
	\caption{\small{The proposed secure ISAC system utilizing an HRIS, intended to enable the localization of both legitimate and malicious users, leading to an improved secrecy rate.}}
	\label{fig: system}
	\end{center}
\end{figure}

Let us consider the wireless system setup depicted in Fig.~\ref{fig: system}, operating within a narrow frequency band, which consists of a multi-antenna BS and an HRIS, intended to enable secure DL communications with a single-antenna UE, while preventing an Eve from intercepting the transmission. For this goal, the HRIS's sensing capabilities are leveraged to localize both legitimate and malicious nodes. The BS is equipped with an $N_{\rm T}$-element Uniform Linear Array (ULA) and supports fully digital TX BF, whereas the HRIS is modeled as a Uniform Planar Array (UPA) with $N_{\rm H} \triangleq N_{\rm RF}N_{\rm E}$ hybrid meta-atoms, where each column of elements is connected to a distinct RX RF chain ($N_{\rm RF}$ in total, with $N_{\rm H}/N_{\rm RF} \in \mathbb{Z}^+_*$), enabling partial signal absorption (determined by the parameter $\varrho\in[0,1]$ to be detailed in the sequel) for baseband processing\cite{alexandropoulos2023hybrid}. For simplicity, we assume $\lambda/2$ spacing between adjacent elements in both the BS and HRIS, where $\lambda = c/f_c$ is the signal wavelength with $c$ denoting the speed of light and $f_c$ is the operating frequency. We also assume that the BS and HRIS know each other's static positions\cite{RIS_challenges}, and that the controller of the metasurface shares its signal observations and phase configurations with a central processing unit hosted at the BS, which is tasked to implement a secure ISAC design, as detailed next in Section~III.

The simultaneous reflection and sensing functionality of the HRIS is controlled through $N_{\rm H}$ identical power splitters \cite{alexandropoulos2023hybrid,alexandropoulos2020hardware,ghazalian2024joint}, which divide the power of the impinging signal at each hybrid meta-atom for the respective operations. For the sensing operation, to feed the absorbed portion of the impinging signal to the $N_{\rm RF}$ RX RF chains, the HRIS applies the analog combining matrix $\W_{\rm H}\in\Compl^{N_{\rm H}\times N_{\rm RF}}$, which is modeled as~\cite{ghazalian2024joint}: $[\W_{\rm H}]_{(l-1)N_{\rm E}+n,j}=w_{l,n}$ for $l=j$ ($l,j=1,\ldots,N_{\rm RF}$) and $[\W_{\rm H}]_{(l-1)M_{\rm E}+n,j}=0$ for $l\neq j$, 
where $|w_{l,n}|=1$ ($n=1,\ldots,N_{\rm H}$) for each non-zero element in $\W_{\rm H}$. In addition, the effective HRIS reflection coefficients are represented by $\boldsymbol{\varphi}\in\Compl^{N_{\rm H}\times 1}$ (i.e., the $1-\varrho$ portion of the impinging signal), assuming $|[\boldsymbol{\varphi}]_n|=1$ $\forall n$. Finally, the BS precodes digitally via the channel-dependent BF vector $\f_{\rm T}\in\Compl^{N_{\rm T}\times 1}$ each complex-valued symbol $s$ such that $\mathbb{E}\{\|\f_{\rm T}s\|^2\}\leq P_{\rm max}$, where $P_{\rm max}$ represents the maximum transmission power.

\subsection{Channel Models}
As shown in the Cartesian coordinate system in Fig.~\ref{fig: system}, the BS is positioned at the origin and the HRIS is placed opposite of it, with its orientation facing the BS and its first element placed at the point $\p_{\rm H}\triangleq[x_{\rm H},y_{\rm H},z_{\rm H}]^{\rm T}\in\mathbb{R}^{3\times1}$. We focus on far-field signal propagation, modeling both the UE and Eve as a point sources positioned between the BS and HRIS, specifically, at the points $\p_{\rm UE}\triangleq[x_{\rm UE},y_{\rm UE},z_{\rm UE}]^{\rm T}\in\mathbb{R}^{3\times 1}$, and $\p_{\rm Eve}\triangleq[x_{\rm Eve},y_{\rm Eve},z_{\rm Eve}]^{\rm T}\in\mathbb{R}^{3\times 1}$ respectively.

The $N_{\rm T}\times 1$ complex-valued DL channel between the BS and either the UE or Eve, as indicated by the index identifier $k\in\{{\rm UE,Eve}\}$, is modeled in the far field as follows:
\begin{align}\label{eq: DL}
    \h_{{\rm DL},k} \triangleq \frac{\lambda}{4\pi\|\p_k\|^2} e^{-\jmath2\pi\frac{\|\p_k\|}{\lambda}}\a_{\rm BS}(\theta_k),
\end{align}
where $\theta_k$ is the azimuth Angle of Departure (AoD), and $\a_{\rm BS}(\cdot)\in\Compl^{N_{\rm T}\times 1}$ is the BS steering vector. 
Similar to \eqref{eq: DL}, the $N_{\rm H}\times N_{\rm T}$ complex-valued Line-of-Sight (LoS) channel between the BS and the HRIS is modeled as follows:
\begin{align}\label{eq: DL_1}
    \H_{\rm BH} \triangleq \frac{\lambda}{4\pi\|\p_{{\rm H}}\|^2}e^{-\jmath2\pi\frac{\|\p_{\rm H}\|}{\lambda}}\a_{\rm H}(\psi_{\rm BR},\phi_{\rm BR})\a_{\rm BS}^{\rm H}(\theta_{\rm BR}),
\end{align}
where $\theta_{\rm BR}$, $\psi_{\rm BR}$, and $\phi_{\rm BR}$ denote respectively the azimuth AoD, as well as the azimuth and elevation Angles of Arrival (AoAs) between the BS and HRIS. In addition, $\a_{\rm H}(\cdot)\in\Compl^{N_{\rm H}\times 1}$ represents the HRIS steering vector. 

The $N_{\rm H}\times N_{\rm T}$ complex-valued channel, which accounts for single-bounce reflections of the BS's signals from the UE and Eve that are received at the HRIS side, can be expressed in the considered far-field regime as follows:
\begin{align}\label{eq: bistatic}
    \H_{{\rm H}}\triangleq \hspace{-0.3cm}\sum_{k\in\{{\rm UE,Eve}\}}\hspace{-0.1cm}a_{{\rm H},k}e^{-\jmath2\pi\frac{d_k}{\lambda}}\a_{\rm H}(\psi_k,\phi_k)\a_{\rm BS}^{\rm H}(\theta_k),
\end{align}
where $\alpha\triangleq\frac{e^{\jmath\omega_k}\lambda^2}{4\pi^{2}\|\p_{k}\|^2\|\p_{\rm H}-\p_{k}\|^2}$ and $d_k\triangleq\|\p_{k}\|+\|\p_{\rm H}-\p_{k}\|$ represent the complex end-to-end path attenuation gain and the distance, respectively, whereas $e^{\jmath\omega_k}$ denotes the complex-valued reflection coefficient with $\omega_k$ being uniformly distributed in $[0,2\pi)$. Finally, the parameters $\psi_k$ and $\phi_k$ correspond to the elevation and azimuth AoAs, respectively.

For the considered system model of Fig.~\ref{fig: system}, the BS and HRIS steering vectors are given respectively by $\a_{\rm BS}(\theta_k)\triangleq\frac{1}{\sqrt{N_{\rm T}}}[1,e^{\jmath2\pi\sin{\theta_k}},\ldots,e^{\jmath(N_{\rm T}-1)\pi\sin{\theta_k}}]^{\rm T}$ and $\a_{\rm H}(\psi_k,\phi_k)\triangleq\a_{\rm H}^{\rm rows}(\phi_k)\otimes\a_{\rm H}^{\rm cols}(\psi_k)$, where $\a_{\rm H}^{\rm rows}(\cdot)\in\Compl^{N_{\rm RF}\times 1}$ and $\a_{\rm H}^{\rm cols}(\cdot)\in\Compl^{N_{\rm E}\times 1}$ are respectively the steering vectors with respect to the azimuth and elevation AoAs, modeled analogously to $\a_{\rm BS}(\cdot)$. According to the this system geometry, the involved AoDs and AoAs in~\eqref{eq: DL}--\eqref{eq: bistatic} can be expressed as:
\begin{align}
    &\nonumber\theta_k = {\rm atan2}( y_k,x_k),\,\phi_k = {\rm atan2}( y_k - y_{\rm H},x_k - x_{\rm H}),\\
    &\nonumber\psi_k = {\rm atan2}\left(z_k - z_{\rm H},r_k\right),\,\phi_{\rm BR}={\rm atan2}\left(z_{\rm H},r_{\rm H} \right),\\
    &\nonumber\theta_{\rm BR}=\psi_{\rm BR}={\rm atan2}(y_{\rm H},x_{\rm H}),
\end{align}
where $r_k\triangleq\|[x_{\rm H} - x_k,y_{\rm H} - y_k]\|$ and $r_{\rm H}\triangleq\|[x_{\rm H},y_{\rm H}]\|$ with ${\rm atan}2(\cdot)$ being the two-argument arctangent function.

\subsection{Received Signal Models and Secrecy Rate}
The baseband received signal at either the UE or Eve can be mathematically expressed as follows:
\begin{equation*}\label{eq:UE_received_signal}
    y_k\!\triangleq\!\left(\h_{{\rm DL},k}\!+\!(1\!-\!\varrho)\h_{{\rm HU},k}{\rm diag}(\boldsymbol{\varphi})\H_{{\rm BR},k}\right)\f_{\rm T}s\!+\!n,
\end{equation*}
where $\h_{{\rm HU},k}$ represents the $1\times N_{\rm H}$ complex-valued channel vector gain between the HRIS and legitimate UE or Eve, which is modeled similar to~\eqref{eq: DL}, and $n\sim\mathcal{CN}(0,\sigma^2)$ models the Additive White Gaussian Noise (AWGN). Recall that the parameter $\varrho\in[0,1]$ is the common power splitting ratio at the HRIS meta-atoms.

Making the reasonable assumption that the static BS-HRIS channel in~\eqref{eq: DL_1} can be accurately estimated and, thus, completely canceled out, the baseband received signal at the outputs of the $N_{\rm RF}$ RX RF chains of the HRIS, considering $T$ information-bearing symbol transmissions, can be mathematically expressed as follows: 
\begin{align}\label{eq: y_ref}
    \Y = [\y(1),\ldots,\y(T)] \triangleq \varrho\W_{\rm H}^{\rm H}\H_{{\rm H}}\f_{\rm T}\s+\N,
\end{align}
where $\s\triangleq[s(1),\ldots,s(T)]\in\Compl^{1\times T}$ and $\N \triangleq [\n(1),\ldots,\n(T)]\in\Compl^{N_{\rm RF}\times T}$, with $\n(t)\sim\mathcal{CN}(0,\sigma^2\I_{N_{\rm RF}})$ $\forall t=1,\dots,T$ being the vector with the transmitted symbols intended for the UE and the matrix with the AWGN vectors per received symbol, respectively.

Capitalizing on the previous signal models, the achievable rates (in bps/Hz) for both the legitimate (i.e., towards the UE) and eavesdropping links can be expressed as follows:
\begin{align*}
    \nonumber\mathcal{R}_{\rm UE}(\f_{\rm T},\boldsymbol{\varphi};\p_{\rm UE}) &\triangleq \log_2\left(1+\sigma^{-2}\|\h_{\rm UE}(\boldsymbol{\varphi},\p_{\rm UE})\f_{\rm T}\|^2\right),\\
    \mathcal{R}_{\rm Eve}(\f_{\rm T},\boldsymbol{\varphi};\p_{\rm Eve}) &\triangleq \log_2\left(1+\sigma^{-2}\|\h_{\rm Eve}(\boldsymbol{\varphi},\p_{\rm Eve})\f_{\rm T}\|^2\right),
\end{align*}
where $\h_{k}(\boldsymbol{\varphi},\p_k)\triangleq\h_{{\rm DL},k}\!+\!(1\!-\!\varrho)\h_{{\rm HU},k}{\rm diag}(\boldsymbol{\varphi})\H_{{\rm BR},k}$. Note that the channels $\h_{{\rm DL},k}$ and $\h_{{\rm HU},k}$ for $k\in\{{\rm UE,Eve}\}$ can be composed using the position $\p_k$'s, as in \eqref{eq: DL}, and the achievable secrecy rate can be computed as the following function of the BS precoder $\f_{\rm T}$ and the HRIS reflection configuration vector $\boldsymbol{\varphi}$ for given UE and Eve 3D positions $\boldsymbol{\eta}\triangleq[\p_{\rm UE}; \p_{\rm Eve}]\in\mathbb{R}^{6\times 1}$: 
\begin{equation}\label{eq: R_s}
\begin{split}
\mathcal{R}_s(\f_{\rm T},\boldsymbol{\varphi};\boldsymbol{\eta})\triangleq\max\big\{0,&\mathcal{R}_{\rm UE}(\f_{\rm T},\boldsymbol{\varphi};\p_{\rm UE})
\\&-\mathcal{R}_{\rm Eve}(\f_{\rm T},\boldsymbol{\varphi};\p_{\rm Eve})\big\}.
\end{split}
\end{equation}

\section{Proposed HRIS-Enabled Secure ISAC}
In this section, we present our communications-centric HRIS-enabled secure ISAC framework, which is optimized to simultaneously estimate both the UE and Eve positions and realize secure DL communications. We begin by deriving the CRB for the UE and Eve positioning parameters, and then present BS and HRIS design frameworks for a genie-aided system with a priori position information as well as a system possessing uncertainty regions for the UE and Eve positions. 

\subsection{PEB Analysis}
It is evident from \eqref{eq: y_ref}'s inspection that, for a coherent channel block involving $T$ unit-powered symbol transmissions $\forall t=1,2,\ldots,T$, yields $T^{-1}\s\s^{\rm H}=1$, hence, the received signal at the outputs of the HRIS's RX RF chains can be modeled as $\y\triangleq{\rm vec}\{\Y\}\sim\mathcal{CN}(\boldsymbol{\mu},\sigma^2\I_{N_{\rm RF}T})$,with mean $\boldsymbol{\mu} \triangleq {\rm vec}\{\varrho\W_{\rm H}^{\rm H}\H_{{\rm H}}\f_{\rm T}\s\}$. The $\J\in\mathbb{R}^{6\times6}$ Fisher Information Matrix (FIM) of $\boldsymbol{\eta}$ can be computed analogously to\cite{kay1993fundamentals} as:
\begin{align}\label{eq: FIM}
    [\J]_{i,j}=\frac{2T\varrho^2}{\sigma^2}\Re\left\{{\rm Tr}\left\{\frac{\partial\Bar{\boldsymbol{\mu}}^{\rm H}}{\partial\eta_i}\frac{\partial\Bar{\boldsymbol{\mu}}}{\partial\eta_j}\right\}\right\}
\end{align}
for $i,j=1,\ldots,6$, where $\Bar{\boldsymbol{\mu}}=\W_{\rm H}^{\rm H}\H_{{\rm H}}\f_{\rm T}$. Focusing on the location-domain parameters $\widetilde{\boldsymbol{\eta}}\triangleq[\theta_{\rm UE},\psi_{\rm UE},\phi_{\rm UE},\theta_{\rm Eve},\psi_{\rm Eve},\phi_{\rm Eve}]\in\mathbb{R}^{6\times 1}$, we now derive the FIM $\widetilde{\J}\in\mathbb{R}^{6\times 6}$ of $\widetilde{\boldsymbol{\eta}}$ as follows \cite[Chapter~3.8]{kay1993fundamentals}:
\begin{align}\label{eq: TPEB}
    \widetilde{\J} = \T^{\rm T}\J\T+\widetilde{\J}_{\rm prior},
\end{align}
where $\T\in\mathbb{R}^{6U\times6U}$ is the transformation matrix expressed as a Jacobian with $[\T]_{i,j}=\partial[\widetilde{\boldsymbol{\eta}}]_i/\partial[{\boldsymbol{\eta}}]_j$, while $\T^{\rm T}\J\T$ and $\widetilde{\J}_{\rm prior}$ respectively represent the FIMs with respect to the observations and, if available, any prior knowledge. Combining all above to quantify sensing performance, we adopt the PEB as our performance metric, computed as:
\begin{align}\label{eq: PEB1}
    {\rm PEB}(\f_{\rm T},\W_{\rm H};\boldsymbol{\eta})=\sqrt{{\rm Tr}\left\{\widetilde{\J}^{-1}\right\}},
\end{align}
which determines the combined positioning accuracy for both the UE and Eve, while their individual localization accuracies can be expressed as follows:
\begin{align}\label{eq: PEB2}
    &\nonumber{\rm PEB}_{\rm UE}(\f_{\rm T},\W_{\rm H};\boldsymbol{\eta})=\sqrt{{\rm Tr}\left\{[\widetilde{\J}^{-1}]_{1:3,1:3}\right\}},\\
    &{\rm PEB}_{\rm Eve}(\f_{\rm T},\W_{\rm H};\boldsymbol{\eta})=\sqrt{{\rm Tr}\left\{[\widetilde{\J}^{-1}]_{4:6,4:6}\right\}}.
\end{align}
Finally, as shown in Appendix, the elements of the FIM in \eqref{eq: FIM}, and consequently, the PEBs in \eqref{eq: PEB1} and \eqref{eq: PEB2} are functions of the BS digital precoding and the HRIS analog combining covariance matrices $\F\triangleq\f_{\rm T}\f_{\rm T}^{\rm H}$ and $\W\triangleq\W_{\rm H}\W_{\rm H}^{\rm H}$. 

\subsection{Genie-Aided Secure ISAC}\label{Frame1}
We aim to jointly optimize the reconfigurable parameters of the BS and HRIS to satisfy a secrecy rate constraint, which will prevent Eve from intercepting DL transmissions, while simultaneously enabling the accurate localization of both the UE and Eve; assuming a data association mechanism, this estimation can be used to formulate the secrecy rate metric. In particular, assuming the availability of some a priori knowledge for the UE and Eve 3D positions, i.e., for $\boldsymbol{\eta}$, we formulate the following design optimization problem:  
\begin{align}
        \mathcal{OP}_1:\nonumber&\underset{\substack{\f_{\rm T},\W_{\rm H},\boldsymbol{\varphi}}}{\min} \,\,{\rm PEB}(\f_{\rm T},\W_{\rm H};\boldsymbol{\eta})\\
        &\nonumber\,\,\text{\text{s}.\text{t}.}\,\mathcal{R}_s(\f_{\rm T},\boldsymbol{\varphi};\boldsymbol{\eta})\geq r_{\rm th},\,\left\|\f_{\rm T}\right\|^2 \leq P_{\rm{\max}},
        \\&\nonumber\,\quad\,|w_{l,n}|=1,\,|[\boldsymbol{\phi}]_n|=1\,\forall l,n.
\end{align}
where $r_{\rm th}$ represents the secrecy rate threshold (in fact, $r_{\rm th}>0$ sufficies), establishing secure DL communications for the legitimate UE. Here, $\mathcal{OP}_1$ is non-convex due to its objective structure and constraints. To address this, we adopt an alternating optimization approach. We first address the non-convexity of the secrecy rate, which is the difference of two concave functions and, thus, not necessarily convex; the HRIS analog combining weight constraints will be incorporated later on. To resolve this, we apply several mathematical simplifications and reformulate the constraint as follows:
\begin{align}
    \nonumber \f_{\rm T}^{\rm H}(\H_{\rm UE}-2^{r_{\rm th}}\H_{\rm Eve})\f_{\rm T}\geq \sigma^2(2^{r_{\rm th}}-1),
\end{align}
where $\H_k\triangleq\h_{{\rm DL},k}^{\rm H}\h_{{\rm DL},k}$ $\forall k\in\{{\rm UE,Eve}\}$ resulting in a common quadratic constraint.

For a given $\W_{\rm H}$ and $\boldsymbol{\varphi}$, $\mathcal{OP}_1$ can be relaxed as follows:
\begin{align}
        \mathcal{OP}_2:\,\,\nonumber&\underset{\substack{\F,\t}}{\min} \,\,\mathbf{1}_{6\times 1}^{\rm T}\t\\
        &\nonumber\text{\text{s}.\text{t}.}\,{\rm Tr}\{(\H_{\rm UE}-2^{r_{\rm th}}\H_{\rm Eve})\F\}\geq\sigma^2(2^{r_{\rm th}}-1),\,\F\succeq0,\\
        &\nonumber\,\quad\begin{bmatrix}
        \widetilde{\J} & \e_a\\
        \e_a^{\rm T} & t_a
        \end{bmatrix}\succeq0\, \forall a=1,\ldots,6,\,{\rm Tr}\{\F\} \leq P_{\rm{\max}},
\end{align}
where $\t\triangleq[t_1,t_2,\ldots,t_6]$ is a set of newly introduced auxiliary variables, $\e_a$ is the $a$th column of a $6\times6$ identity matrix, and all elements of $\widetilde{\J}$ have been reformulated as in the Appendix. Assuming SemiDefinite Relaxation (SDR), and by relaxing the rank-one constraint of $\F\triangleq\f_{\rm T}\f_{\rm T}^{\rm H}$, $\mathcal{OP}_2$ becomes a convex SemiDefinite Program (SDP), which can be efficiently solved using standard convex solvers\footnote{The optimized precoding vector can be directly derived as $\f_{\rm opt}=\u_1\sqrt{\rho_1}$, where $\u_1$ is the principal singular vector and $\rho_1$ is the corresponding singular value of $\F$. Note that the optimized solution satisfies the rank-one constraint.} (e.g., CVX). 

For a given $\f_{\rm T}$ and $\boldsymbol{\varphi}$, $\mathcal{OP}_1$ can be simplified as follows:
\begin{align}
        \mathcal{OP}_3:\,\,\nonumber&\underset{\substack{\{\W_l\}_{l=1}^{N_{\rm RF}},\t}}{\min} \,\,\mathbf{1}_{6\times 1}^{\rm T}\t\\
        &\nonumber\text{\text{s}.\text{t}.}\,{\rm diag}(\W_l)=1\,\W_l\succeq 0,\,\forall l,\\
        &\,\quad\begin{bmatrix}
        \widetilde{\J} & \e_a\\
        \e_a^{\rm T} & t_a
        \end{bmatrix}\succeq0\, \forall a=1,\ldots,6,
\end{align}
where $\W_{\rm H}\W_{\rm H}^{\rm H}$ forms a block diagonal matrix: $\W_l\triangleq\W_{\rm H}(k_l:lN_{\rm E},l)\W_{\rm H}^{\rm H}(k_l:lN_{\rm E},l)$ $\forall l$ with $k_l\triangleq(l-1)N_{\rm E}+1$, and $\W_{\rm H}\W_{\rm H}^{\rm H}=\sum_{l=1}^{N_{\rm RF}}\W_l$. Each element of the FIM is reformulated accordingly to incorporate the block diagonal structure of $\W_l$ $\forall l$. Additionally, each rank-one constraint is replaced with a positive semidefinite one utilizing SDR. As a result, $\mathcal{OP}_3$ can be efficiently solved similar to $\mathcal{OP}_2$, with $\W_{\rm H}(k_l:lN_{\rm E},l)$ $\forall l$ constructed from the phase shifts of $\W_l$'s principal singular vector.

The HRIS reflection configuration vector can be finally designed via the following optimization problem that uses the definition $\boldsymbol{\varphi}=e^{\jmath\boldsymbol{\upsilon}}$ with the vector $\boldsymbol{\upsilon}$ including the $N_{\rm H}$ tunable reflection coefficients:
\begin{align}
        \mathcal{OP}_4:
        &\,\nonumber\underset{\substack{\boldsymbol{\upsilon}}}{\max} \,\,\mathcal{R}_s\left(\f_{\rm T},e^{\jmath\boldsymbol{\upsilon}};\boldsymbol{\eta}\right)\\
        &\nonumber\,\,\text{\text{s}.\text{t}.}\,\, -\frac{\pi}{2}\leq[\boldsymbol{\upsilon}]_n\leq\frac{\pi}{2}\,\forall n=1,2,\ldots,N_{\rm H},
\end{align}
where $\mathcal{R}_s(\f_{\rm T},e^{\jmath\boldsymbol{\upsilon}};\boldsymbol{\eta})$ is the secrecy rate, as defined in~\eqref{eq: R_s}, evaluated for the reflection coefficients $e^{\jmath\boldsymbol{\upsilon}}$. This problem can be restructured as a standard quadratic fractional programming problem with a unit modulus constraint. It can be efficiently solved using existing optimization methods for conventional RISs~\cite{Tsinghua_RIS_Tutorial}, or the approaches presented in \cite{shen2019secrecy}.

Putting all above together, $\mathcal{OP}_1$ is iteratively solved by sequentially computing $\f_{\rm T}$, $\W_{\rm H}$, and $\boldsymbol{\varphi}$ as the solutions of the problems $\mathcal{OP}_2,\mathcal{OP}_3$, and $\mathcal{OP}_4$, respectively, until a convergence criterion is satisfied, or a maximum number of iteration has been reached.

\subsection{Secure ISAC with Uncertainty Regions}\label{Frame2}
In practice, it is more reasonable to design the proposed HRIS-enabled secure
ISAC system considering candidate areas where the UE and Eve can be located, since a priori knowledge about their positions can be unavailable. To this end, we assume, in this section, that UE's and Eve's 3D position vectors $\p_{\rm UE}$ and $\p_{\rm Eve}$ lie respectively within the uncertainty regions $\mathcal{K}_{\rm UE}$ and $\mathcal{K}_{\rm Eve}$. Those regions can be disjoint or sharing common points, though hereinafter, we discretize each of them into $K$ equidistant points each, namely, $\mathcal{K}_{\rm UE}=\{\p_{{\rm UE},b_1}\}_{b_1=1}^K$ and $\mathcal{K}_{\rm Eve}=\{\p_{{\rm Eve},b_2}\}_{b_2=1}^K$, respectively, having disjoint elements. Following this modeling of the UE and Eve uncertainty regions, we now focus on the following secure ISAC optimization problem for the design of the reconfigurable parameters of the BS and HRIS:
\begin{align}
        \mathcal{OP}_5:\nonumber&\underset{\substack{\f_{\rm T},\W_{\rm H},\boldsymbol{\varphi}}}{\min} \,\,\underset{\boldsymbol{\eta}\in\mathcal{H}}{\max}\,\,{\rm PEB}(\f_{\rm T},\W_{\rm H};\boldsymbol{\eta})\\
        &\nonumber\,\,\text{\text{s}.\text{t}.}\,\,\mathcal{R}_{\rm UE}^{\rm min}(\f_{\rm T},\boldsymbol{\varphi})-\mathcal{R}_{\rm Eve}^{\rm max}(\f_{\rm T},\boldsymbol{\varphi})\geq r_{\rm th},
        \\&\nonumber\quad\,\,\left\|\f_{\rm T}\right\|^2 \leq P_{\rm{\max}},\,|w_{l,n}|=1,\,|[\boldsymbol{\phi}]_n|=1\,\forall l,n,
\end{align}
where $\mathcal{H}\triangleq \mathcal{K}_{\rm UE}\cup \mathcal{K}_{\rm Eve}$ includes all UE and Eve position points for the PEB evaluation, while $\mathcal{R}_{\rm UE}^{\rm min}(\f_{\rm T},\boldsymbol{\varphi})$ and $\mathcal{R}_{\rm Eve}^{\rm max}(\f_{\rm T},\boldsymbol{\varphi})$ represent respectively the minimum UE and the maximum Eve achievable rates within the respective discretized uncertainty regions, which are obtained as follows:
\begin{align}
     &\nonumber\mathcal{R}_{\rm UE}^{\rm min}(\f_{\rm T},\boldsymbol{\varphi}) \triangleq \underset{\{\p_{{\rm UE},b_1}\}_{b_1=1}^K}{\min}\,\mathcal{R}_{\rm UE}\left(\f_{\rm T},\boldsymbol{\varphi};\p_{{\rm UE},b_1}\right),\\
     &\nonumber\mathcal{R}_{\rm Eve}^{\rm max}(\f_{\rm T},\boldsymbol{\varphi}) \triangleq \underset{\{\p_{{\rm Eve},b_2}\}_{b_2=1}^K}{\min}\,\mathcal{R}_{\rm Eve}\left(\f_{\rm T},\boldsymbol{\varphi};\p_{{\rm Eve},b_2}\right).
\end{align}

To solve $\mathcal{OP}_5$, we introduce the auxiliary variables $t$ and $\varepsilon_{a,b}$'s, with $a=1,\ldots,6$ and $b=1,\ldots,K^2$, to reformulate it in an epigraph form, as follows~\cite{chierchia2015epigraphical}:
\begin{align}
        \mathcal{OP}_6:\nonumber&\underset{\substack{\f_{\rm T},\W_{\rm H},\boldsymbol{\varphi},\{\varepsilon_{a,b}\}_{\forall a,b},t}}{\min} \,\,t\\
        &\nonumber\text{\text{s}.\text{t}.}\,\begin{bmatrix}
        \widetilde{\J}(\F,\W;\boldsymbol{\eta}_b) & \e_a\\
        \e_a^{\rm T} & \varepsilon_{a,b}
        \end{bmatrix}\succeq 0\,\forall b=1,\ldots,K^2,
        \\&\nonumber\,\quad \varepsilon_{a,1}+\ldots+\varepsilon_{a,K^2}\leq t\,\forall a=1,\ldots,6,
        \\&\nonumber\,\quad \mathcal{R}_{\rm UE}^{\rm min}-\mathcal{R}_{\rm Eve}^{\rm max}\geq r_{\rm th},\, {\rm Tr}\{\F\} \leq P_{\rm{\max}},
        \\&\nonumber\,\quad|w_{l,n}|=1,\,|[\boldsymbol{\phi}]_n|=1\,\forall l,n,
\end{align}
where $\widetilde{\J}(\F,\W;\boldsymbol{\eta}_b)$ indicates the FIM in~\eqref{eq: TPEB} with its elements reformulated as detailed in the Appendix. In particular, this FIM expression is evaluated at all UE and Eve 3D positions $\boldsymbol{\eta}_b\triangleq[\p_{{\rm UE},b_1};\p_{{\rm Eve},b_2}]$ $\forall b_1,b_2=1,\ldots,K$. Note that $\mathcal{OP}_6$ remains non-convex for the same reasons that $\mathcal{OP}_1$ is non-convex. To address this, we apply the same alternating optimization strategy used for treating $\mathcal{OP}_1$, where now instead of solving $\mathcal{OP}_1$ directly, we iteratively solve $\mathcal{OP}_2$ through $\mathcal{OP}_4$. Using similar simplifications, we can reformulate $\mathcal{OP}_6$ as a convex SDP with respect to the unknown design variables $\F$ and $\W$, and as a fractional programming problem for the $\boldsymbol{\varphi}$ design.

 \begin{figure*}[!t]
  \begin{subfigure}[t]{0.48\textwidth}
  \centering
    \includegraphics[width=\textwidth]{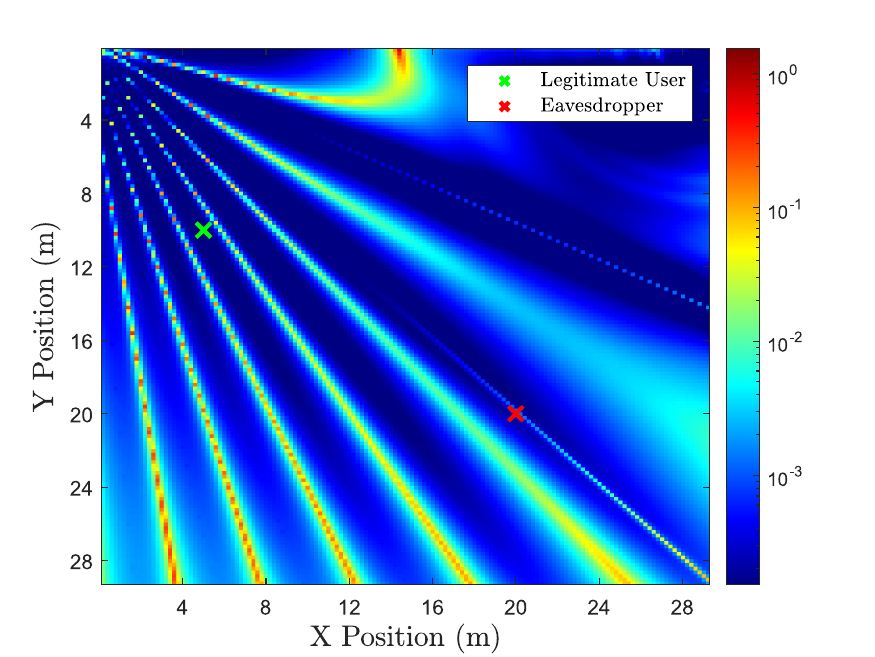}
  \end{subfigure}\hfill
  \begin{subfigure}[t]{0.48\textwidth}
  \centering
    \includegraphics[width=\textwidth]{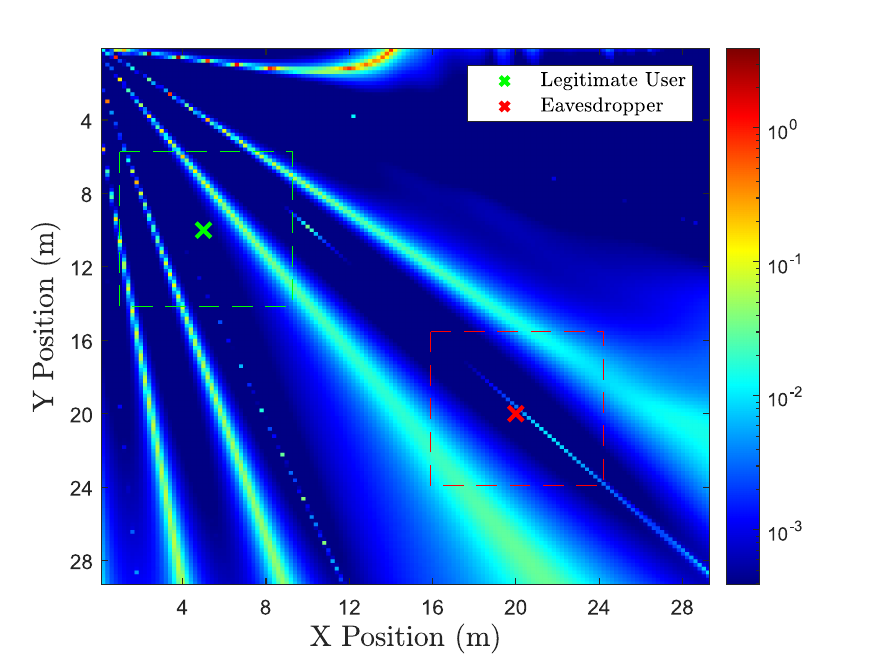}
  \end{subfigure}
  \caption{\small{Area-wide PEB for $P_{\rm \max}=20$ dBm and $r_{\rm th}=7$ dB, considering UE and Eve placed at the points $[5, 10, 2]$ and $[20, 20, 2]$. Left: A priori position information available; and Right: Knowledge of uncertainty regions (dashed squares), spaced at $4$ meters apart from the center in the $x$- and $y$-direction.
  }}
  \label{fig:PEB}
\end{figure*}

\subsection{Complexity Analysis}
In SDPs, the primary bottleneck in computational complexity often arises from the presence of numerous Linear Matrix Inequalities (LMIs). It is clear that $\mathcal{OP}_2$, $\mathcal{OP}_3$, and particularly $\mathcal{OP}_6$, involve a significant number of LMIs, with their complexity scaling with the number of antenna elements $N_{\rm T}$ at the BS, the number of RIS elements $N_{\rm H}$, and the number of points within $\mathcal{K}_{\rm UE}$ and $\mathcal{K}_{\rm Eve}$. Utilizing interior-point methods for the SDPs, with a worst-case time complexity of $\mathcal{O}(n^2\sum_{i=1}^C m_i^2 + n\sum_{i=1}^C m_i^3)$~\cite{nemirovski2004interior}, where $n$ denotes the number of optimization variables, $C$ represents the number of LMI constraints, and $m_i$ is the column or row size of each $i$th LMI constraint. Specifically, for $\mathcal{OP}_2$, we have $n = N_{\rm T}^2+6$, $C = 7$, and $m_i = 7$ for $i<C$, while $m_i = N_{\rm T}$ for $i=C$. Assuming $N_{\rm T}^2\gg\sum_{i=1}^{C-1}m_i^2$, a condition that is typically satisfied in most deployment scenarios, the resulting approximate complexity is $\mathcal{O}(N_{\rm T}^6)$. Similarly, for $\mathcal{OP}_3$, we have $n=N_{\rm RF}N_{\rm E}^2+6$, $C=6+N_{\rm RF}$, and $m_i=7$ for $i<6$ and $m_i=N_E$ for $i>6$. Making a similar assumption as in $\mathcal{OP}_2$, the approximate complexity becomes $\mathcal{O}(N_{\rm RF}^3N_{\rm E}^6)$. Additionally, $\mathcal{OP}4$ can be efficiently solved in $\mathcal{O}(N_{\rm H}^{3.5})$ time complexity. Taking all above terms into account, the overall per-iteration time complexity is given by:
\begin{align}
    \mathcal{O}(N_{\rm T}^6+N_{\rm RF}^3N_{\rm E}^6+N_{\rm H}^{3.5}).
\end{align}
By following the same steps for the complexity analysis of $\mathcal{OP}_6$, we can approximate the worst-case per-iteration complexity, assuming $K^2\leq N_{\rm T}\leq N_{\rm H}$, as follows:
\begin{align}
    \mathcal{O}(N_{\rm T}^8+N_{\rm RF}^4 N_{\rm E}^{8}+N_{\rm H}^{3.5}).
\end{align}
It is worth noting that the aforedescribed optimization analysis directly solves the considered SDPs without employing dimensionality reduction or any lower-bound approximations. Although techniques, such as heuristic codebook designs \cite{keskin2022optimal} and PEB lower-bound approximations \cite{gavras2025near}, can significantly reduce computational complexity by largely avoiding LMIs, their integration is left for the journal version of this work.

\section{Numerical Results and Discussion}
In this section, we evaluate the performance of the proposed communications-centric HRIS-enabled secure ISAC system. We have conducted $500$ Monte Carlo simulations under a narrowband system setup centered at $20$~GHz, with each coherent channel block comprising $T=200$  transmissions. Unless otherwise specified, both the UE and Eve were randomly positioned within a 3D area bounded by $[0,30]$ meters along the $x$- and $y$-axes, centered at $z=2$ at the $z$-axis. The HRIS was fixed at the location $\p_{\rm RIS}=[0,30,5]$. We have considered a BS equipped with a ULA of $N_{\rm T}=16$ antennas, an HRIS featuring $N_{\rm RF} = 4$ RX RF chains each connected to $N_{\rm E} = 8$ hybrid meta-atoms, and AWGN's with variance $\sigma^2 = -100$ dBm.

In Fig.~\ref{fig:PEB}, the PEB performance over a 2D geographic area between the BS and the HRIS for both the genie-aided (left subfigure) and the proposed (right subfigure) secure ISAC schemes is illustrated, considering the BS transmit power at $P_{\rm \max}=20$ dBm, a secrecy rate threshold of $r_{\rm th}=0$ bps/Hz, and fixed positions for the UE and Eve at the 3D points $[5, 10, 2]$ and $[20, 20, 2]$, respectively. For the proposed scheme, we have simulated two square-shaped uncertainty regions, one for the UE and the other for Eve, extending $4$ meters around each of them in both the $x$- and $y$-direction, as shown in the dashed areas in the right subfigure. Each uncertainty region was discretized in $K=7$ equidistant points. It can be observed from both subfigures that effective localization of both the UE and Eve can be achieved, while meeting the secrecy rate constraint. This happens for the case in the left subfigure where a priori knowledge of the UE and Eve is available, but also on the right subfigure, where satisfactory PEB performance is exhibited at the exact UE and Eve positions as well as in the points around them and within the uncertainty regions.
In fact, by considering multiple points within each uncertainty region, the proposed ISAC scheme allocates more resources for sensing, resulting in wider beams that effectively cover the entire area, as opposed to the narrow beams focusing on single positions with the genie-aided scheme. It thus becomes evident that optimizing with respect to multiple closely spaced positions within each uncertainty region can significantly enhance sensing performance, as compared to other state-of-the-art schemes that focus on some a priori position knowledge~\cite{alexandropoulos2025extremely}, which can be imperfect.

\begin{figure}[!t]
\centering
\includegraphics[width=\columnwidth]{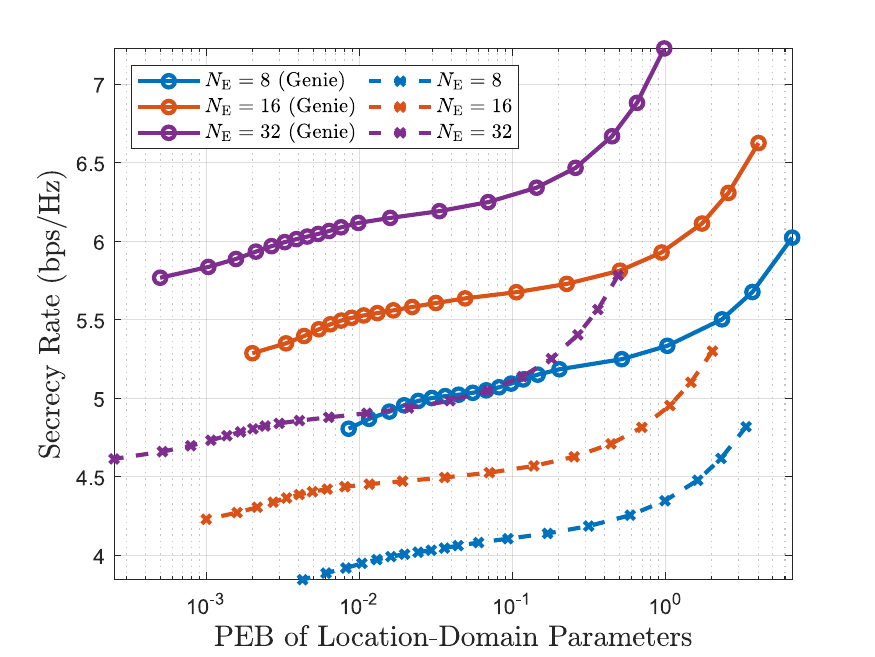}
\caption{\small{Secrecy rate vs. sensing performance for both the genie-aided and proposed schemes w.r.t $\rho\in[0,1]$, considering $P_{\rm \max}=20$ dBm and $N_{\rm E}=\{8,16,32\}$ hybrid meta-atoms at the HRIS.}
}
\label{fig:Tradeoff}
\end{figure}

Figure~\ref{fig:Tradeoff} illustrates the tradeoff between localization accuracy and secrecy rate performance for a system setup with $P_{\rm \max}=20$ dBm, comparing both the genie-aided and proposed ISAC designs for different numbers of hybrid meta-atoms at the HRIS, in particular, $N_{\rm E}=\{8,16,32\}$. We have, specifically, reformulated the secrecy-rate-constrained problems $\mathcal{OP}_{1}$ and $\mathcal{OP}_{6}$ as dual-objective optimization problems, with the tradeoff being controlled by the absorption coefficient $\rho$, in line with the approach presented in \cite{alexandropoulos2025extremely}. For the uncertainty regions of the proposed scheme, we have followed the same configuration with the right subfigure of Fig.~\ref{fig:PEB}. As expected, higher absorption by the HRIS enhances sensing performance, while increased power portion devoted to the reflection operation favors improved secrecy rates. It is, furthermore, shown that, as the number of HRIS elements increases, the genie-aided scheme achieves superior secrecy rate performance compared to the proposed robust secure ISAC design. However, the former exhibits poorer localization accuracy. This aligns with the observations in Fig.~\ref{fig:PEB}'s right subfigure, where it can be seen that, optimizing over multiple points within an uncertainty region, leads to better positioning performance. As previously discussed, this is attributed to the larger amount of system resources being directed towards the sensing operation.

\section{Conclusion}
This paper proposed a communications-centric secure ISAC framework capitalizing on the dual-functionality capability of HRISs. The proposed joint BS precoding and HRIS configuration design establishes simultaneous estimation of both the UE and Eve positions with secure DL communications towards the UE. A CRB analysis was presented to evaluate the positioning accuracy of both users, which was deployed in the objective of the proposed secure ISAC problem formulation including a secrecy rate constraint. Our numerical investigations showcased the simultaneous secure communications and sensing capability of the proposed scheme, highlighting the role of the HRIS absorption coefficient on the trade-off between those two operations.


\section*{Appendix}\label{App}
Let the definitions $\F\triangleq\f_{\rm T}\f_{\rm T}^{\rm H}$ and $\W\triangleq\W_{\rm H}\W_{\rm H}^{\rm H}$. Then, each $(i,j)$th element ($i,j=1,\ldots,6$) of the FIM in \eqref{eq: FIM} can be described as follows:
\begin{align}
    \nonumber[\J]_{i,j}&=\frac{2T\varrho^2}{\sigma^2}\Re\left\{{\rm Tr}\left\{\f_{\rm T}^{\rm H}\frac{\partial\H_{{\rm H}}^{\rm H}}{\partial\eta_i}\W_{\rm H}\W_{\rm H}^{\rm H}\frac{\partial\H_{{\rm H}}}{\partial\eta_j}\f_{\rm T}\right\}\right\}\\
    &\nonumber=\frac{2T\varrho^2}{\sigma^2}\Re\left\{{\rm Tr}\left\{\frac{\partial\H_{{\rm H}}^{\rm H}}{\partial\eta_i}\W_{\rm H}\W_{\rm H}^{\rm H}\frac{\partial\H_{{\rm H}}}{\partial\eta_j}\f_{\rm T}\f_{\rm T}^{\rm H}\right\}\right\}\\
    &=\frac{2T\varrho^2}{\sigma^2}\Re\left\{{\rm Tr}\left\{\frac{\partial\H_{{\rm H}}^{\rm H}}{\partial\eta_i}\W\frac{\partial\H_{{\rm H}}}{\partial\eta_j}\F\right\}\right\},
\end{align}
indicating that the FIM depends on the $\F$ and $\W$ matrices.
\bibliographystyle{IEEEtran}
\bibliography{ms}

\end{document}

%% file: Definitions.tex
\renewcommand{\a}{\mathbf{a}}

\newcommand{\e}{\mathbf{e}}
\newcommand{\f}{\mathbf{f}}

\newcommand{\h}{\mathbf{h}}

\newcommand{\n}{\mathbf{n}}

\newcommand{\p}{\mathbf{p}}

\newcommand{\s}{\mathbf{s}}
\renewcommand{\t}{\mathbf{t}}
\renewcommand{\u}{\mathbf{u}}

\newcommand{\y}{\mathbf{y}}

\newcommand{\F}{\mathbf{F}}

\renewcommand{\H}{\mathbf{H}}
\newcommand{\I}{\mathbf{I}}
\newcommand{\J}{\mathbf{J}}

\newcommand{\N}{\mathbf{N}}

\newcommand{\T}{\mathbf{T}}

\newcommand{\W}{\mathbf{W}}

\newcommand{\Y}{\mathbf{Y}}

\newcommand{\Compl}{\mbox{$\mathbb{C}$}}

\renewcommand{\Re}{\mathrm{Re}}